\documentclass{ws-ijmpd}

\newcommand{\dm}[2]{(m^2_{#1}-m^2_{#2})}

\begin{document}

\markboth{George W.S. Hou} {CPV for BAU}

\catchline{}{}{}{}{}
%
%%%%%%%%%%%%%%%%%%%%%%%%%%%%%%%%%%%%%%%%%%%%%%%%%%%%%%%%%%%%%%%%%%%%

\title{SOURCE OF CP VIOLATION FOR THE BARYON ASYMMETRY OF THE UNIVERSE
% \footnote{For the title, try
%not to use more than 3 lines. Typeset the title in 10~pt Times
%roman, uppercase and boldface.}
}

\author{GEORGE W.S. HOU%\footnote{Typeset names in
%8~pt roman, uppercase. Use the footnote to indicate the
%present or permanent address of the author.}
}

\address{Department of Physics, National Taiwan University\\
Taipei, Taiwan 10617
%\footnote{State completely without
%abbreviations, the affiliation and mailing address, including
%country. Typeset in
%8~pt Times italic.}
\\
wshou@phys.ntu.edu.tw
\\
\phantom{ l} and
\\
\phantom{ l} National Center for Theoretical Sciences, National
Taiwan University, \phantom{ l}\\
Taipei, Taiwan 10617}

%\author{SECOND AUTHOR}

%\address{}

\maketitle

\begin{history}
\received{Day Month Year}
\revised{Day Month Year}
\comby{Managing Editor}
\end{history}

\begin{abstract}
We give a description of why the existence of a fourth generation
is likely to provide enough CP violation for baryogenesis, and
trace how this observation came about. We survey the current
experimental and theoretical pursuits and outline a research
agenda, touching upon unitarity violation and very heavy chiral
quarks, and comment on how the electroweak phase transition
picture might be altered.
\end{abstract}


\section{Introduction}

It was the great physicist Andrei Sakharov who made the link
between the puzzling experimental discovery of CP violation (CPV),
with the even more puzzling Baryon Asymmetry of the Universe
(BAU): the absence of antimatter from the observable Universe. The
BAU puzzle is as follows. At the Big Bang, equal amounts of matter
and antimatter ought to be produced. Of course, they will mutually
re-annihilate as the Universe cools, and indeed this feeds eventually
the Cosmic Microwave Background radiation. But why then is {\it any}
matter left, and at roughly $10^{-9}$ of the primordial production?
Sakharov's three conditions\cite{Sakharov} for this to occur is:
\begin{romanlist}[(ii)]
 \item Baryon Number Violation;
 \item CP Violation;
 \item Deviation from Equilibrium.
\end{romanlist}

It is truly remarkable that the Standard Model (SM) satisfies
condition (i) in a nontrivial way, provides CPV phase(s) in the
charged current through quark mixing, and one is hopeful for
nonequilibrium through the ``condensation" that lead to spontaneous
electroweak symmetry breaking (EWSB). Alas, the SM seems
\emph{insufficient} in conditions (ii) and (iii): the amount of
CPV in the three generation Kobayashi--Maskawa model falls
far short from what is needed, as we shall see in the next section,
while the phase transition seems too smooth because the Higgs boson
is not light enough.

It has therefore been popular to invoke ``Baryogenesis through
Leptogenesis", namely that BAU occurs first through
lepton--antilepton imbalance, then transferred to baryons by the
electroweak forces in SM. I offer some comments. Leptogenesis
based on traditional seesaw mechanism for generating tiny neutrino
mass, through right-handed Majorana scale at $10^{12}$ GeV or
higher, is rather beautiful. However, it appears to be
``metaphysics", in the sense that it can not be experimentally
tested in the foreseeable future! Then, there are the Type II and
III, etc. seesaw models that bring in more assumptions, in good
part to make them more accessible at the LHC (or future machine),
and the models become less beautiful.

This pushes the traditional-minded physicists like myself to yearn
for the SM, since it satisfies all {\it necessary} conditions of
Sakharov, albeit insufficiently in two of them. With the caution
that we have no right that ``the theory of our time" would touch
so deeply the core to the Universe (and Our Existence), we do like
to ask:

\vskip0.2cm
 \centerline{\it Can one restore TeV Scale Baryogenesis?}

\vskip0.2cm
 \centerline{\it What about the Source of CP Violation?}

\vskip0.2cm
 This talk tries to touch upon these profound issues,
especially on the CPV front.

\section{Tracing a Thread in the Tapestry: \ CPV on Earth}

CP violation was forced upon us by experimental discovery, which
caused the pure minds such as Dirac to depress. But it in fact
opened our minds further to deeper truths on the antimatter world
that Dirac himself uncovered for us.

\subsection{Experimental knowledge of CPV}

Sakharov wrote down his conditions in 1966 (published in 1967),
which was clearly stimulated by the experimental
observation,\cite{CPV1964} in 1964, of CP violation in
$K_2 \to \pi^+\pi^-$ decay, now interpreted as the physical
$K_L^0$ meson having a small admixture of the $K_1$ state.
The 1980 Nobel prize was awarded to James Cronin and Val Fitch
for their experimental discovery. The pursuit was on for the
``direct" CPV (DCPV), i.e. in decay, within the kaon system,
which was finally established\cite{PDG} in 1999.

It was the two (then) young Japanese physicists, Makoto Kobayashi
and Toshihide Maskawa (KM), who pointed out\cite{KM} in 1972
(published in 1973) that if a third generation (3G) of quarks exist,
then a unique CPV phase appeared in the $3\times 3$ quark mixing matrix
governing the charged current. It is remarkable that, at that time,
even two generations were not completely established. But within a
few years, the $c$ quark, the $\tau$ lepton, and the $b$ quark
were all discovered, although it took another 18 years before the
top quark was discovered at the Tevatron.

But the main issue for KM was CP violation. The picture was
convincingly confirmed\cite{PDG} between the Belle and BaBar
experiments in 2001, and the
pair was awarded 1/2 the 2008 Nobel prize. What is remarkable, and
reflecting the prowess of these B factory experiments, is that
DCPV in the $B$ system, in the form of difference in rate for
$B^0 \to K^-\pi^+$ vs. $\bar B^0 \to K^+\pi^-$, was the
highlight observation of 1994. It came a mere three years after
the Nobel prize defining measurement of mixing-dependent CPV
(TCPV) in 2001, which is in contrast to the tortuous path of
35 years for the kaon system.
We will discuss further developments which sprang from the
observation of DCPV in the B system, in the next section.

\subsection{KM model and its limitations}

\subsubsection{Complex dynamics: KM sector of SM}

What KM pointed out was that, while the $2\times 2$ quark mixing
matrix of the charged current (weak coupling $g$ modulated as
$gV_{ij}$) is real, a unique, irremovable phase appeared in the
$3\times 3$ generalization. The unitary matrix $V$ can
be parameterized\cite{PDG} in the form where the $2\times 2$ sector
is real to very good approximation, while it is traditional (a phase
convention) to put the unique CPV phase in the $V_{ub}$ element,
which is then reflected in the $V_{td}$ element by unitarity, or
$VV^\dag = V^\dag V = I$.

Unitarity of $V$ correlates multiple physical measurables
involving flavor and CPV. One such condition is the relation
\begin{equation}
V_{ud} V_{ub}^* + V_{cd} V_{cb}^* + V_{td} V_{tb}^* = 0\,,
\label{eqn1}
\end{equation}
from $\{VV^\dag\}_{db} = 0$. The KM condition for CPV is that the
triangle formed by Eq.~(\ref{eqn1}) should be {\it nontrivial},
i.e. the {\it area} $A$ of the triangle should not vanish.
Remarkably, while many relations, or triangles, similar to
Eq.~(\ref{eqn1}) can be written or formed, they all have the same
area $A$, reflecting the unique CPV phase in the 3G KM model.

For Eq.~(\ref{eqn1}) to be {\it nontrivial}, the sides of the
triangle should not be colinear. It was measuring the finite angle
between $V_{td} V_{tb}^*$ and $V_{cd} V_{cb}^*$ (the latter
defined real in standard\cite{PDG} convention) in 2001, together with
knowledge of the strength of the sides $V_{ud} V_{ub}^*$ and
$V_{cd} V_{cb}^*$ as well as many other flavor/CPV observables,
that confirmed the nontrivial realization of Eq.~(\ref{eqn1}), hence
the CPV phase of the KM model.

\subsubsection{Jarlskog invariant and CPV}

Besides the nontrivial realization of Eq.~(\ref{eqn1}), a further
subtlety can be inferred from the original KM argument: all
like-charged quark pairs must be nondegenerate in mass! Otherwise,
if their is just one pair of, say $d$ and $s$ quarks, that are
degenerate in mass, then one finds a phase freedom that can absorb
the single CPV phase, and effectively one is back to the two
generation case with vanishing CPV.

An algebraic construction, known as the Jarlskog invariant,\cite{Jarlskog85} nicely summarizes the nontrivialness of Eq.~(\ref{eqn1}) and the
nondegeneracy requirement:
\begin{equation}
J = \dm{t}{u} \dm{t}{c} \dm{c}{u}
    \dm{b}{d} \dm{b}{s} \dm{s}{d} \, A\,,
 \label{eqn2}
\end{equation}
where $A$ is the triangle area as defined before, while the
appearance of every (like-charged) pair mass difference ensures
that $J$ would vanish with the degeneracy.

$J$ in Eq.~(\ref{eqn2}) is not merely a transcription of the
wording of previous prerequisites, but has powerful algebraic
roots. It can be derived from $J \equiv {\rm
Im}\,\det\left[m_um_u^\dag,\ m_dm_d^\dag\right]$ for the case of 3G.
Thus, in terms of the Jarlskog invariant, one has

\vskip0.2cm
 \centerline{CPV \ {\it iff} \ $\,J \neq 0$.}

\subsubsection{The ``Lore" for insufficient BAU from CPV in KM}

In his Nobel lecture, Kobayashi admitted that ``Matter dominance
of the Universe seems requiring new source of CP violation",\cite{Nobel}
i.e. beyond the 3G model he and Maskawa presented.
In fact, it is known\cite{Peskin} that $J$ seems short by at least $10^{-10}$! Let me give\cite{Hou09} a heuristic, dimensional argument for why this is so.

The issue of BAU is not so much the disappearance of antimatter,
i.e. the apparent $n_{\overline{\cal B}}/n_\gamma \cong 0$, but
that {\it some}, in fact a tiny amount of matter remain (which
contains {\it us}!!), namely $n_{{\cal B}}/n_\gamma = (6.2 \pm
0.2) \times 10^{-10}$ as measured by WMAP, which is the $10^{-9}$
quoted earlier in the Introduction. Thus, the actual
\emph{asymmetry}, or BAU, is 100\%, but the challenge is to
explain $n_{{\cal B}}/n_\gamma \neq 0$, and to account for the
tiny amount. Note that this is a dimensionless number, while $J$ of
Eq.~(\ref{eqn2}), the source of CPV, carries 12 mass dimensions.
Normalizing by $T \sim 100$ TeV, the electroweak phase transition
temperature (equivalently one could normalize by the v.e.v.), then
inserting all quark masses gives $J/T^{12} \sim 10^{-20}$, which
can now be compared with $n_{{\cal B}}/n_\gamma \sim 10^{-9}$.
This is the ``Lore" that the CPV in KM model is too small by
at least 10 billion.

Further inspection of Eq.~(\ref{eqn2}) shows that $A\sim 3\times
10^{-5}$ as measured, though small, is not the major culprit. The
real issue is that quark masses (except $m_t$) are too small: the
powers of $m_s^2 \, m_c^2 \, m_b^4$ as compared to $T^8$ are just
too small!

\section{Soaring to the Heavens: \ 3G $\to$ 4G}

The way the previous section ended has already planted the seed
for the main observation of this section. But let us trace through
the way it actually came about. In effect, it arose from broadening of
the Mind by \emph{Nature} writing.

\subsection{The Thread again}

Experiment is our modern age Delphi \emph{oracle}, and what it
utters sometimes has more than one interpretations.

The Thread that lead was the hint, at 2.4$\sigma$ level for Belle,\cite{belle04}
that emerged with the 2004 observation of DCPV in $B^0 \to
K^+\pi^-$: the asymmetry $A_{K^+\pi^0}$ for the analogous charged
$B^\pm$ meson decays seemed different from $A_{K^+\pi^-}$ for
neutral $B$ meson decays. With similar effect seen by BaBar, the
plenary speaker at ICHEP 2004 from Belle, Yoshi Sakai, questioned\cite{Sakai}
whether this hinted at large electroweak (or $Z^0$) penguin, hence
implied New Physics. The point is that a virtual $Z^0$ could
convert to a $\pi^0$, but not a charged pion, hence the $Z^0$
penguin contributes to $B^\pm \to K^\pm \pi^0$, but is less
effective for $B^0 \to K^\pm\pi^\mp$. But if $P_{\rm EW}$ is the
culprit, then it must arise from New Physics, as there is
vanishing CPV phase in $b\to s$ penguin transitions within SM, as
it is governed by $V_{ts}V_{tb}^*$, which is effectively real for
3G.

Shocked while writing the first draft of this Belle paper --- the
counterintuitive difference was never predicted --- it reminded me
of my first B paper,\cite{HWS87} which was on the related
electroweak penguin process $b \to s \ell^+\ell^-$ (the
$\ell^+\ell^-$ takes the place of the $\pi^0$). Prior to that
paper, $G_F$ power counting had lead people to discard the $Z^0$
penguin as compared to the photonic penguin. At $G_F^2$ order, the
former should be small compared with the latter, which is at
$\alpha G_F$ order. Or so it seems: since $G_F$ has $-2$ mass
dimension, there should be some $m^2$ to make the comparison with
the photonic penguin. One would again dismiss it by taking $m \sim
m_b$ naively. However, it turns out that $m \sim m_t$, the top
quark in the loop that could be heavy.

Direct computation showed that for large $m_t$ ($\gtrsim 2M_W$),
the $b \to s \ell^+\ell^-$ rate grew almost like $m_t^2$, and the
heavy quark effect is {\it nondecoupled}. We should be familiar
with the usual decoupling theorem, where heavy masses are
decoupled from scattering amplitudes, such as in QED and QCD,
since they only appear in propagators. However,
\emph{nondecoupling} appears because Yukawa couplings $\lambda_Q
\propto m_Q/v$, where $v$ is the v.e.v., appear in the numerator
and can counteract the propagator damping. Thus, the nondecoupling
phenomena is a \emph{dynamical} effect, and is a subtlety of
spontaneously broken {\it chiral} gauge theories.

My first B paper turned out to be also my first four generation
(4G) paper, where the nondecoupling effect of 4G $t'$ quark could
be easily more prominent. So, I went ahead and demonstrated with
two associates the efficacy of the 4G $t'$ quark, that it
could\cite{HNS} drive apart $A_{K^+\pi^0}$ from $A_{K^+\pi^-}$,
for a range of parameters in $m_{t'}$ and $V_{t's}^*V_{t'b} \equiv
r_{sb} \, e^{i\phi_{sb}}$. As a corollary, since the $Z^0$ penguin
and the box diagram are cousins of each other, the CPV effect of
nondecoupling of $t'$ in $b\to s$ $Z^0$ penguin should have
implications for CPV in the $\bar B_s$--$B_s$ mixing via the box
diagram, which we will discuss in the next section.

\subsection{Nature writing}

Because direct CPV, including the DCPV difference $\Delta A_{K\pi}
\equiv A_{K^+\pi^0} - A_{K^+\pi^-}$, are simple ``bean counts",
the Belle experiment decided to write a paper for the journal
\emph{Nature} to highlight the effect. With even CDF joining the
measurement, the asymmetry $A_{K^+\pi^-}$ became firmly
established around $-10\%$. Therefore, although $A_{K^+\pi^0}$ was
not yet firmly established, the unanticipated $\Delta A_{K\pi}$,
measured now by a single experiment (Belle) to
be\cite{belleDeltaA} $+0.164 \pm 0.037$ with 4.4$\sigma$
significance, is {\it very large}: the difference is larger than
the already impressively large $A_{K^+\pi^-} \simeq -10\%$ (cf.
$|\varepsilon'_K| \sim 10^{-6}$).

Although ``the oracle spoke", the effect put forward by this paper
was not widely accepted as indicating New Physics. Perhaps the
particle physics community treat \emph{Nature} announcements no
better than the \emph{New York Times}. There was also the issue
that large $\Delta A_{K\pi}$ could be interpreted as an enhanced
color-suppressed tree amplitude $C$ that has a considerable strong
phase difference with $T$, the regular tree amplitude. But the
actual ``\emph{Nature} writing", in ``explaining CPV to
biologists", got me ``out of my mind", which I turn to in the next
subsection.

\subsection{Providence}

The heuristic, dimensional analysis argument for why the KM
mechanism for CPV falls far short of BAU makes clear that the
culprit is the smallness of lighter quark masses. As we tried to
convey to the editor of \emph{Nature} the relevance of large
$\Delta A_{K\pi}$ to readers of their journal, one day late summer
2007, it occurred to me that, if there is 4G and one shifts by one
generation in Eq.~(\ref{eqn2}) for the Jarlskog invariant $J$
(recall that one needs 3 generation for the KM mechanism of CPV,
hence one is discarding the first generation for a 2-3-4 world),
one gets
\begin{equation}
J_{(2,3,4)}^{sb} = \dm{t'}{c} \dm{t'}{t} \dm{t}{c}
           \dm{b'}{s} \dm{b'}{b} \dm{b}{s} A_{234}^{sb},
 \label{eqn3}
\end{equation}
where one sees that, besides $m_t^4 \to m_{t'}^2 (m_{t'}^2 -
m_t^2)$, the extreme suppression factor of $m_s^2 \, m_c^2 \,
m_b^4$ is replaced by $m_b^2 \, m_t^2 \, m_{b'}^4$. Even for
$A_{234}^{sb}$ comparable in strength to $A$ (numerical analysis
of $\Delta A_{K\pi}$ suggested\cite{HNS} a factor of 30), this
lead to a gain of $10^{13}$--$10^{15}$ for $J_{(2,3,4)}^{sb}$ over
$J$ for 3G, for $m_{b'},\ m_{t'} \in (300,\ 600)$ GeV, and clearly
removes the verdict that KM mechanism for CPV falls far short of
observed BAU.

The fact that now one seems to have enough CPV within SM, at the
cost of 3G $\to$ 4G, makes one wonder whether \emph{Mother Nature}
might actually use this? True enough, it was amusing to receive
the arXiv number of ``.1234", a sure sign of Providence, when I
posted the paper from a Zurich hotel room in March 2008, before
heading for Moriond. But then, a probability of $10^{-3}$ is
nothing compared to the gain of a thousand trillion ($10^{15}$).
As an anecdote, the paper eventually appeared in the \emph{Chinese
Journal of Physics} (published in Taiwan) in 2009.

%\begin{figure}[pb]
%\centerline{\psfig{file=ijmpdf1.eps,width=4.7cm}} \vspace*{8pt}
%\caption{A schematic illustration of dissociative recombination.
%The direct mechanism, 4m$^2_\pi$ is initiated when the molecular
%ion S$_{\rm L}$ captures an electron with kinetic energy.
%\label{f1}}
%\end{figure}

\section{2007--2010: \ 4G Rehab}

The stiffness one faced on 4G studies were not without reason: the
fourth generation had become rather exotic with data from LEP. For
the detailed early numeric study of 4G effect on $\Delta
A_{K\pi}$, I was lucky to publish two papers in \emph{Phys. Rev.
Lett.} The first one in 2005 may be because it was a timely
response to some emergent phenomenon from the B factories. For the
second,\cite{HLMN07} applying PQCD factorization at
next-to-leading order (NLO), may be due to its sheer technicality
\ldots

But one could clearly feel the rehabilitation of 4G during the
year of 2010, perhaps even becoming a mild fashion. Such was not
the situation back in 2007.

\subsection{Why not 4G? }

Let me use the words of an experimentalist, Alison
Lister\cite{Lister} of CDF (and now ATLAS), at the ICHEP 2010
conference. Why not four generations? There are\cite{PDG} two
issues:
\begin{itemize}
\item $Z$-width measurement from LEP: perfect fit with only three
light neutrinos;
\item Electroweak effects: $S$, $T$ fits (severely) constrain
available 4G phase space.
\end{itemize}

For the first, traditional, fourth generation show stopper, Lister
counters that the true constraint is $m_{\nu_4} > M_Z/2$. Let me
add to that, by first changing the notation of the possible new
fourth neutral lepton and denote it as $N^0$, to avoid the
connotation of lightness that comes with ``$\nu_4$". It should be
emphasized that, since the discovery of atmospheric neutrino
oscillations in 1998,\cite{PDG} we know that neutrinos have mass,
implying the existence of another mass scale. This logically
refutes the traditional strict interpretation that a fourth light
neutrino is excluded by LEP data. It is indeed excluded, but we
already know there is New Physics in the neutrino or neutral
lepton sector. We then stress that the neutral lepton $N^0$ is
very hard to access in the near future at the LHC (or through
neutrino oscillations), unless it is of Majorana nature with
v.e.v. scale masses.

The second problem of electroweak (EW) $S$ and $T$ constraints are
potentially more serious. But, as pointed out by Kribs, Plehn,
Spannowsky and Tait in 2007,\cite{Kribs} these constraints have
been over-interpreted (by PDG): 4G is in fact allowed by EW
radiative corrections, and one could even argue that sometimes it
gives better agreement together with a \emph{heavier} Higgs boson.
This has been further followed up by Chanowitz,\cite{Chanowitz}
and with the response\cite{Erler} from Erler and Langacker not
fully refuting, it is a main cause of the mini-revival of 4G in
the past few years.

\subsection{Touching more Earth: CPV in $B_s$ system}

There are other reasons for the gradual move to more favorable
view (as compared to the past) on 4G, arising from flavor and
especially CPV studies of the $B_s$ system. I am fond of quoting
the CDF citation\cite{CDF08} of myself ``George Hou predicted the
presence of a $t'$ quark with mass \ldots\ to explain the Belle
results and predicted {\it a priori} the observation of a large
$CP$-violating phase in $B_s \to J/\psi \, \phi$ decays".The
wording ``predicted {\it a priori}" is especially amusing, and
should be a reminder to theorists. In any case, this refers to my
work on the corollary of large $\sin2\Phi_{B_s}$ for the $t'$
quark interpretation of the $\Delta A_{K\pi}$ ``anomaly".

We showed in PQCD at LO in 2005,\cite{HNS} then at NLO in
2007,\cite{HLMN07} that 4G can in principle generate $\Delta
A_{K\pi}$. The prediction in 2005 was that $\sin2\Phi_{B_s}$,
defined as the CP phase of the $b\bar s \to s\bar b$ box diagram
(mediating $\bar B_s \to B_s$, similarly defined as
$\sin2\Phi_{B_d} \equiv \sin2\phi_1 \equiv \sin2\beta$ for $\bar
B_d \to B_d$), would be in the range of $-0.2$ to $-0.7$. This was
refined\cite{HNS07} in 2007 to $-0.5$ to $-0.7$ after $\Delta
m_{B_s}$ was observed by the CDF experiment in 2006. The reason
that CDF jovially quoted me in summer 2008 is because three
consecutive measurements at the Tevatron ($\sin2\beta_s \equiv
-\sin2\Phi_{B_s}$ for CDF, and $\sin\phi_s \equiv \sin2\Phi_{B_s}$
for D$\emptyset$) gave large central values. The combined
significance, however, had dropped to 2.1$\sigma$ by summer
2009.\cite{Punzi}

My 2005 and 2007 studies were based on $m_{t'} =$ 300 GeV. As the
mass bounds were rising, I was working with an associate on a 500
GeV update. The experimental developments in 2010 were actually
mixed, but also turned up the heat.
First, it was the D$\emptyset$ announcement\cite{DzeroASL} in May
of significant $a_{sl}^s$ (something akin to the $\epsilon_K$ but
for the $B_s$ system). This strengthened the indication of
deviation from SM (i.e. 3G). I had commented,\cite{HM07} with an
associate, on the previous round of D$\emptyset$ studies, and had
mentioned that 4G could lead to a sizable $a_{sl}^s$. With the new
result, which gives the same central value but improves the errors
by a factor of two, I did not want to write another paper. But I
placed a comment in a conference talk,\cite{Top10} that the new
D$\emptyset$ result, if true, would violate a bound already
stressed in Ref.~\refcite{HM07}, hence probably suggests
hadronically enhanced (i.e. OPE violating) $\Delta\Gamma_s$
values. Then came the CDF result\cite{betas10} on $\sin2\beta_s$
that was less discrepant with SM, implying a smaller value. In the
meantime, and prior to the CDF update, I had pointed
out\cite{HouMa10} that the expected value for $\sin2\Phi_{B_s}$
was weaker (nominally $-0.3$) for the heavier $m_{t'} = 500$ GeV
case.

So, the fourth generation ``prediction" is still robust, but would
now need LHCb to verify. It must have been rather sad for the B
workers at CDF (who had remeasured $\Delta m_{B_s}$) when they
opened the box for 5.2 fb$^{-1}$ data. Had the added data firmed
up the 2008--2009 indication, it would allow the possible capture
of $\sin2\Phi_{B_s}$ at the ``evidence" or better level with the
full Tevatron dataset, hence would have constituted a New Physics
discovery. With the low central value, and with already half the
dataset of Run II used, there is no hope for any future claim to
``evidence", and the torch is thereby passed to LHCb. On the
theory front, the papers\cite{4G10} by Soni and associates, Buras
and associates, and Lenz and associates in the first half of 2010,
together with other studies, clearly ushered in the ``4G
rehabilitation".

\subsection{The Pursuit, and its dilemma/opportunity}

In retrospect, actually much if not most highlights of flavor and
CPV physics were learned through the \emph{nondecoupling} effect:
the GIM mechanism, the charm mass, $\varepsilon_K$ from the $s\bar
d \to d\bar s$ box; heavy top as inferred from large $B_d$ mixing
($b\bar d \to d\bar b$ box), with the consequent CPV phase
measurement, and the small $\varepsilon'/\varepsilon$ due to $s\to
d$ $Z$-penguin and $Z$-penguin enhanced $b\to s\ell^+\ell^-$ rate.
\emph{All from boxes and $Z$ penguins!} If there is 4G, we already
saw the possible effect on $B_s$ system. Other measurables to
watch would be $A_{\rm FB}(B\to K^*\ell^+\ell^-)$, redux of
$\sin2\phi_1/\beta$ and $\varepsilon_K$, $Z\to b\bar b$, maybe
$\sin2\Phi_D$, and especially $K_L\to \pi^-\nu\nu$ (KOTO
experiment). That is, an agenda for all aspects of flavor physics
and CP violation, all as a consequence of large Yukawa couplings!

But, nothing can replace direct search for the 4G $t'$ and $b'$
quarks, and we are on the verge of transit from the Tevatron to
the LHC era.

The pursuit at the Tevatron has been vigorous, with the mass bound
ever rising. The current CDF limit is\cite{Lister}
\begin{equation}
 m_{t'} > 335\ {\rm GeV, \ \ 95\%\ C.L.},
 \label{eqn4}
\end{equation}
based on 4.6 fb$^{-1}$ data. But a persistent irritation since
earlier analyses with smaller datasets, is the weakening of the
bound from what was expected, due to excess events at high $M_{\rm
reco}$ (reconstructed mass) and $H_T$ (a scalar sum of transverse
energies). With D$\emptyset$ now observing something
similar\cite{Lister} but giving a weaker bound, it is not clear
whether the excess events are due to common misunderstanding of
background, or something genuine. CDF has pursued the much cleaner
signature of same-sign dileptons via $b'\bar b'$ pair production,
followed by $b'\to tW$ decay, reaching mass bounds
similar\cite{CDFb'} to Eq.~(\ref{eqn4}) for ${b'}$.

With the successful 2010 run of LHC at 7 TeV, the table is turning
to the ATLAS and CMS experiments. Hereby lies both a dilemma, as
well as an opportunity. With just 1 fb$^{-1}$ data, the bound on
4G masses at LHC would reach beyond 500 GeV,\cite{CMS7TeV} which
is roughly the unitarity bound\cite{CFH79} where
\emph{perturbative} partial wave unitarity, or probability
conservation, breaks down. How does one continue the pursuit? With
the available energy at the LHC, clearly one should not stop
searching at 500 GeV. Besides the need for theoretical guidance
for continued search, precisely because perturbation theory would
breakdown, one comes face to face with some rather interesting
issues related to \emph{strong} Yukawa couplings, the origin of
the aforementioned nondecoupling.

The most tantalizing conjecture is:
 \vskip0.15cm
 \centerline{\underline{Could EWSB be due to $b'$ and $t'$ near or above
  the unitarity bound?}}
 \vskip0.1cm
A conjecture, traced to Nambu (the recipient of the other half of
2008 Nobel prize), is that perhaps $\bar QQ$ could develop a
v.e.v., i.e. condense, by large Yukawa coupling!(?) To seriously
address these issues, one needs a nonperturbative platform of
study, and the only one we know, is on the lattice. A study of the
strong Higgs-Yukawa sector on the lattice has therefore been
initiated.

\subsection{The ``3 + I" approach  ---  a research agenda}

Without further ado, let me outline an approved five-year
(starting August 2010) research program, what I dubbed the
``\textbf{3} + \textbf{I}" approach under the title of
``Beyond Kobayashi-Maskawa
--- Towards Discovery of 4th Generation Quarks at the LHC".

The ``\textbf{3}" is a three-pronged approach to the associated
physics. Naturally, there is the direct search with the CMS
detector. We have also purposefully built up a new theory group,
both for the LHC era in general, and to provide phenomenology
support for the experimental effort. The third arm is a consortium
of Taiwan and DESY-Zeuthen on the aforementioned strong
Higgs-Yukawa on the lattice. Note that results from the lattice
study would become desperately needed to pursue beyond the
expected 2011 data, which would touch and could reach beyond the
unitarity bound.

This approved five-year project has loftier experimental goals: to
uplift the past platform into the full, long term run plan of the
LHC. As such, one needed to expand beyond the Taiwan CMS
contribution to the Preshower subdetector during the past decade.
We were lucky to become part of the CERN/Taiwan center, one of the
three (the other two are PSI/ETHZ and DESY/Aachen/Karlsruhe)
centers for module production for the CMS Pixel Upgrade Phase I,
targeted for completion in 2016. Such ``deeper" involvement within
CMS longer term hardware effort will certainly assure our longer
term physics program.

\section{Conclusion: \ Know in 3--5(--7) Years}

The most important point of this talk is Eq.~(\ref{eqn3}), where
the 1000 trillion ($10^{13}$--$10^{15}$ for $m_{b'},\ m_{t'}$
ranging from 300 to 600 GeV, with $A_{234}^{sb}$ not less in
strength than $A$) gain in CPV over 3G, hence likely enough CPV
for BAU. It makes one suspect that \emph{maybe there is a fourth
generation!} We have discussed flavor/CPV aspects of, as well as
direct search for, 4G quarks. The Tevatron should still be
watched, but clearly the mantle has passed to the LHC:
\begin{itemize}
 \item $\sin2\Phi_{B_s}$ ``Confirmation" --- should be ``easy" at LHCb;
 \item $b'$ and $t'$ Discovery --- straightforward, and able to cover full terrain,
\end{itemize}
except for unitarity bound issues for the latter.

Within 3 to 5 years, maybe 7, we should know the answer. That is
one advantage of 4G vs. other New Physics scenarios (e.g. related
to BAU). And if we find the answer in the affirmative, we may have
brought down Heaven on Earth, namely that we might attain
realistic understanding of BAU, from ``the theory of our time".

Within a matter of years, direct search at the LHC for heavy $b'$
and $t'$ quarks would have hit the unitarity bound. How
\emph{Nature} cures this perturbative malady may shed light on the
source of electroweak symmetry breaking, and the existence and
nature of the Higgs boson. That would be a huge bonus to the 4G
program.

\section{Postscript: What about the Strength of Phase Transition? }

One may perceive a remaining obstacle for electroweak
baryogenesis, even if 4G is established, i.e. condition (iii) of
Sakharov, or departure from equilibrium.
In the standard Higgs potential approach, the strength of phase
transition is controlled by the cubic term in the Higgs field. For
$m_H > 72$ GeV, which seems the case experimentally, Kajantie,
Laine, Rummukainen and Shaposhnikov\cite{KLRS96} have done a
lattice study to show that the transition is only a crossover.
With 4G and without any new bosons, it is still insufficient. The
basic reason is that the cubic term receives only bosonic
contributions, and the $W$ and $Z$ in SM are too light. The remedy
is therefore to put in more bosons, such as light stop in
supersymmetric framework.

I mention some caveats.
First, the ``Nambu $\bar QQ$ pairing", or condensation due to
strong Yukawa, should affect the cubic term.
Second, the (multi-)Higgs field(s) would be likely composite with
strong Yukawa couplings.
Finally, the standard approach treats the Higgs as elementary,
i.e. structureless.  Composite Higgs, which has not been seriously
studied for phase transitions at finite temperature, would change
the scenario.

Could the \emph{nonperturbative} Yukawa couplings of 4G quarks
save the day? This is another issue to be studied by the lattice
Higgs-Yukawa effort.

\section*{Acknowledgments}

We are grateful to the National Science Council of Taiwan for the
approval of the Academic Summit program that supports the future
studies mentioned in this talk.

%\section{References}

%\begin{thebibliography}{000} %for 3 digits
%\begin{thebibliography}{00}  %for 2 digits


\begin{thebibliography}{0}    %for 1 digit

%%journal paper

\bibitem{Sakharov}
 A.D. Sakharov,
 \emph{Pis'ma~Zh.~Eksp.~Teor.~Fiz.} \textbf{5} (1967) 32
  [\emph{JETP~Lett.} \textbf{5} (1967) 24].

\bibitem{CPV1964}
 J.H.~Christenson, J.W.~Cronin, V.L.~Fitch, and R.~Turlay,
 \emph{Phys.~Rev.~Lett.} \textbf{13} (1964) 138.

\bibitem{KM}
 M. Kobayashi and T. Maskawa,
 \emph{Prog.~Theor.~Phys.} \textbf{49} (1973) 652.

\bibitem{PDG}
 K. Nakamura \emph{et al.} (Particle Data Group),
  \emph{J.~Phys.~G} \textbf{37} (2010) 075021.

\bibitem{Jarlskog85}
  C.\ Jarlskog,
  \emph{Phys.~Rev.~Lett.} {\bf 55} (1985) 1039;
  \emph{Z.~Phys.~C} {\bf 29} (1985) 491.

\bibitem{Nobel}
 See http://nobelprize.org/nobel\_prizes/physics/.

\bibitem{Peskin}
 See, e.g. M.E. Peskin,
  %Song of the Electroweak Penguin.
  {\it Nature}\ {\bf 452} (2008) 293.

\bibitem{Hou09}
  W.-S.~Hou,
  \emph{Chin.~J.~Phys.} {\bf 47} (2009) 134
   [arXiv:0803.1234 [hep-ph]].

%%collaboration

\bibitem{belle04}
 Belle Collab. (Y. Chao, P. Chang {\it et al.}),
  %Evidence for Direct $CP$ Violation in $B^0\to K^+\pi^-$ Decays.
  {\it Phys.~Rev.~Lett.} {\bf 93} (2004) 191802.

\bibitem{Sakai}
  Y.~Sakai,
  %``Recent results on B decays,''
  \emph{Int.~J.~ Mod.~Phys. A} {\bf 20} (2005) 5059.

\bibitem{HWS87}
  W.-S.~Hou, R.S.~Willey and A.~Soni,
  {\it Phys.\ Rev.\ Lett.}\ {\bf 58} (1987) 1608.

\bibitem{HNS}
  W.-S.~Hou, M.~Nagashima and A.~Soddu,
  {\it Phys.\ Rev.\ Lett.}\  {\bf 95} (2005) 141601.

\bibitem{belleDeltaA}
 Belle Collab. (S.-W. Lin, Y. Unno, W.-S. Hou, P. Chang {\it et al.}),
  %Difference in direct $CP$ violation between charged and neutral
  %$B$ meson decays.
  {\it Nature}\ {\bf 452} (2008) 332.
%
\bibitem{HLMN07}
 W.-S. Hou, H-n. Li, S. Mishima, and M. Nagashima,
   %Fourth generation $CP$ violation effect on $B\to K\pi, \phi K$
   %and $\rho K$ in next-to-leading-order perturbative  QCD.
   {\it Phys.\ Rev.\ Lett.}\ {\bf 98} (2007) 131801.

\bibitem{Lister}
 A. Lister (on behalf of the CDF and D$\emptyset$ Collaborations),
 Search for fourth generation $t'$ quark at the Tevatron,
 to appear in a special issue of \emph{Proceedings of Science (PoS)} for
 the \emph{35th Int. Conf. on High Energy Physics}, Paris, 2010.

\bibitem{Kribs}
  G.D. Kribs, T. Plehn, M.S. Spannowsky and T.M.P. Tait,
  \emph{Phys. Rev.} D {\bf 76} (2007) 075016.

\bibitem{Chanowitz}
  M.~Chanowitz,
  \emph{Phys. Rev. D} {\bf 79} (2009) 113008.

\bibitem{Erler}
  J.~Erler and P.~Langacker,
  %``Precision Constraints on Extra Fermion Generations,''
  \emph{Phys.\ Rev.\ Lett.}\  {\bf 105} (2010) 031801.

\bibitem{CDF08}
 CDF Collab. public note CDF/ANAL/BOTTOM/PUBLIC/9458,
 August 7, 2008.

\bibitem{HNS07}
  W.-S.~Hou, M.~Nagashima, A.~Soddu,
  \emph{Phys.\ Rev.}\ D {\bf 76} (2007) 016004.

%%proceedings
\bibitem{Punzi}
 G. Punzi, Flavour physics at the Tevatron,
 in {\it Proc. 2009 Europhysics Conference on High Energy Physics},
 PoS E {\bf PS-HEP2009}, 022 (2009).

\bibitem{DzeroASL}
 D$\emptyset$ Collab. (V.M. Abazov \emph{et al.}),
  \emph{Phys.\ Rev.\ Lett.}\  {\bf 105} (2010) 081801.

\bibitem{HM07}
  W.-S.~Hou and N.~Mahajan,
  %``Remarks on the A*(SL) and $\Delta$ Gamma($s$) studies at the Tevatron and
  %beyond,''
  \emph{Phys.\ Rev.\  D} {\bf 75} (2007) 077501.

\bibitem{Top10}
 G.W.-S. Hou, Fourth generation: towards effect of large Yukawa coupling,
 invited plenary talk at \emph{3rd International Workshop on Top
 Quark Physics} (TOP 2010), Bruges, Belgium, 31 May - 4 Jun 2010,
 arXiv:1007.2288 [hep-ph].

\bibitem{betas10}
 CDF Collab. public note CDF/ANAL/BOTTOM/PUBLIC/10206, November 2,
 2010.

\bibitem{HouMa10}
  W.-S.~Hou and C.-Y.~Ma,
  \emph{Phys.\ Rev.\  D} {\bf 82} (2010) 036002.

\bibitem{4G10}
   A.~Soni, A.K.~Alok, A.~Giri, R.~Mohanta and S.~Nandi,
   %``SM with four generations: Selected implications for rare B and K decays,''
   \emph{Phys.\ Rev.}\  D {\bf 82} (2010) 033009;
   A.J.~Buras, B.~Duling, T.~Feldmann, T.~Heidsieck, C.~Promberger and S.~Recksiegel,
   %``Patterns of Flavour Violation in the Presence of a Fourth Generation of
   %Quarks and Leptons,''
   \emph{JHEP} {\bf 1009} (2010) 106;
   O.~Eberhardt, A.~Lenz and J.~Rohrwild,
   %``Less space for a new family of fermions,''
   \emph{Phys.\ Rev.\  D} {\bf 82} (2010) 095006.

\bibitem{CDFb'}
 CDF Collab. (T. Aaltonen \emph{et al.}),
  \emph{Phys.\ Rev.\ Lett.}\  {\bf 104} (2010) 091801.

\bibitem{CMS7TeV}
   CMS Collab., {The CMS physics reach for searches at 7 TeV}, CMS NOTE-2010/008.

\bibitem{CFH79}
  M.S.~Chanowitz, M.A.~Furman and I.~Hinchliffe,
  %``Weak Interactions Of Ultraheavy Fermions. 2,''
  \emph{Nucl.\ Phys.\  B} {\bf 153} (1979) 402.
%
\bibitem{KLRS96}
  K.~Kajantie, M.~Laine, K.~Rummukainen and M.E.~Shaposhnikov,
  %``Is there a hot electroweak phase transition at m(H) > approx. m(W)?,''
  \emph{Phys.\ Rev.\ Lett.}\  {\bf 77} (1996) 2887.



\end{thebibliography}
\end{document}